\documentclass[superscriptaddress, aps, pra,twocolumn,showpacs,floatfix]{revtex4-2}
\usepackage[utf8]{inputenc}
\usepackage{graphicx,amsmath,amsfonts,amssymb}
\usepackage{color}
\usepackage[colorlinks=true, allcolors={blue}]{hyperref}
\usepackage{graphicx}
\usepackage{epstopdf}
\usepackage{float}
\usepackage{placeins}
\usepackage{comment}

\newcommand{\orcid}[1]{\href{https://orcid.org/#1}{\includegraphics[width=7pt]{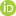}}}

\begin{document}

\title{Two-stroke thermal machine using spin squeezing operation}

\author{Carlos H. S. Vieira\orcid{0000-0001-7809-6215}}
\email{carloshsv09@gmail.com}
\affiliation{Center for Natural and Human Sciences, Federal University of ABC, Avenida dos Estados 5001, 09210-580, Santo Andr\'{e}, S\~{a}o Paulo, Brazil.}
\affiliation{Departamento de F\'isica, Universidade Federal do Piau\'i, Campus Ministro Petr\^onio Portela, 64049-550, Teresina, PI, Brazil.}

\author{Jonas F. G. Santos\orcid{0000-0001-9377-6526}}
\email{jonassantos@ufgd.edu.br}
\affiliation{Faculdade de Ci\^{e}ncias Exatas e Tecnologia, Universidade Federal da Grande Dourados,
Caixa Postal 364, Dourados, CEP 79804-970, MS, Brazil.}

\begin{abstract}
Quantum thermal machines are powerful platforms to investigate how quantum effects impact the energy flow between different systems. We here investigate a two-stroke cycle in which spin squeezing effects are intrinsically switched on during all the operation time. By using Kitagawa and Ueda’s parameter and the $\ell_1$-norm to compute the degree of spin squeezing and the quantum coherence, we first show that the greater the spin squeezing effect, the greater the amount of coherence in the energy basis. Then, we investigate the engine performance given the amount of spin squeezing into the system. Our results show that even assuming an always-on spin squeezing, which is directly associated with the amount of entropy production in the cycle, it is possible to find a better set of efficiency and extracted power for the engine provided a high level of control over the relevant parameters, i.e., the operation time and the squeezing intensity. 

\end{abstract}

\maketitle

\section{Introduction}

The recent advances in quantum technologies have demonstrated that the energetic and entropic aspects are important and useful to optimize quantum protocols~\cite{Auffeves2022}. Energy dissipation is associated with how fast a given protocol is implemented and the entropy production is related to the irreversibility of such a protocol \cite{Landi2021}. Despite these questions appeared in classical thermodynamics, as systems are miniaturized into the quantum regime, interesting effects start to affect the thermodynamics of protocols, such as quantum coherence and quantum correlations \cite{Kosloff2013, Adesso2018, Goold,Deffner2019}. In this regime, quantum fluctuations are as ubiquitous as thermal fluctuations, and this can be employed to boost quantum protocols beyond their classical counterparts as well as modify the proper description of energetics exchange~\cite{Campisi_RMP11,Esposito_RMO09}. This effervescent branch of physics is known as quantum thermodynamics and its impact on future devices and energetic exchanges is evidenced by a series of groundbreaking theoretical and experimental contributions~\cite{Myers2022, vieira_jmr23,Anders2016,serra2014,Chu2022}.

Just as classical thermal machines were relevant to establishing the grounds for thermodynamics and the first industrial revolution, the study of quantum thermal machines has been used to form a set of solid knowledge about quantum thermodynamics and move forward into further quantum revolutions. The literature on quantum thermal machines is vast, and here we highlight that most studies concern the quantum version of the Otto cycle~\cite{Norton_pre24,Nori_pre07}, and different quantum resources to boost the cycle performance, as quantum correlations \cite{Xiao2023, Altintas2014, Asadian2022} and quantum coherence \cite{Camati2019, Peterson2019, Lin2021}. It is well-known that irrespective of the quantum aspects of the working substance, in most cases, the thermal baths and their dynamics follow the Born-Markovian approximation. The maximum efficiency in this setup is the Otto limit, with the power output depending on the speed of the cycle and on the driven Hamiltonian structure \cite{Camati2019}. To circumvent this fact, many authors have proposed the use of structured thermal baths, with the most famous being the squeezing thermal bath \cite{Manzano2018, Klaers217, Assis2020, Assis2021}, resulting in the celebrated generalized Carnot efficiency \cite{Rossnagel2014, Abah2014}. Other models also have been employed, for instance, using $\mathcal{PT}$-symmetric Hamiltonians \cite{Santos2021, Santos2023}, quantum measurements \cite{lisboa2022experimental, Bresque2021, Campisi2019, Ding2018, Jordan2020}, non-Markovian baths \cite{Camati2020, Ptaszynski2022, Cavaliere2022}, and correlated thermal baths \cite{Chiara2020}. 

Despite the relevance of the quantum Otto cycle, it is not the only possibility, with studies focusing on two and three strokes thermal machines \cite{Santos2023, lisboa2022experimental, Talkner2017, Elouard2017, Brandner2015, Mohammady2017}. A two-stroke quantum thermal machine comprises one process involving the thermalization of each component with local thermal baths and one unitary interaction between the parts. The subsystems are general and can be spins or quantum harmonic oscillators \cite{Timpanaro2019, Herrera2023, Sacchi2021}. Recently, a two-stroke quantum thermal machine where the interaction is tailored to perform a total or partial SWAP operation has been used to analyze the work statistics and thermodynamics uncertainty relations (TURs)~\cite{Timpanaro2019,Sacchi2021} and implement a kind of heat engine that allows to achieve an efficiency above the standard Carnot limit~\cite{Herrera2023}.

In this work, we consider a finite-time two-stroke thermal machine fueled by two spins. The interaction between the spins is mediated by employing the one-axis twisting nonlinear spin-squeezing interaction coupled with an external transversal field. This structure for the model allows us to obtain important results concerning the amount of squeezing in the cycle and its performance as well as it is experimentally doable, for instance, in Nuclear Magnetic Resonance (NMR) employing a liquid sample of a quadrupolar system~\cite{Auccaise_prl15} or a system of two-mode Bose-Einstein Condensate (BEC)~\cite{Guang_prl07}. The spin squeezing is quantified in terms of Kitagawa and Ueda's parameter and it is related to the amount of quantum correlations (coherence) employing the $l_1$-norm and all nonequilibrium thermodynamics quantities are analytically obtained. Our results indicate that by having fine-tuned control over the parameters along the cycle, it is possible to reach a good performance even when operating in a finite-time regime. Despite previous works considering two-stroke cycles \cite{Timpanaro2019, Sacchi2021}, in these cases a SWAP operation performs the interaction stroke, whereas in our case the spin squeezing mediates the interaction. This is one of the main goals in our model, i.e., how to engender a two-stroke machine where the spin-squeezing interaction is always on. Besides, we explicitly consider the coherence along the cycle and how it is related to the amount of squeezing.

The present work is organized as follows. Section \ref{secII} introduces some basic concepts, such as a nonlinear spin interaction and how to quantify spin squeezing employing Kitagawa and Ueda parameter. Section \ref{secIII} describes the two-stroke thermal machine based on nonlinear spin squeezing operation and its operation regimes. Section~\ref{secV} outlines our results. Finally, in Section~\ref{secVI} we summarized our remarks and conclusions.


\section{Nonlinear interaction and spin squeezing}\label{secII}

Recently, different aspects of spin squeezing have been extensively examined both in terms of theory and experiment with relevant applications in entanglement~\cite{Sorensen_nature01,Toth_prep09,Amico_RMP08}, improvement of measurement precision~\cite{Wineland_pra92,Pezze_prl09}, optical atomic clocks~\cite{Robinson_NatPhy24}, and also in important areas such as quantum simulations~\cite{Kaubruegger} and quantum computation~\cite{Lu_2023}. Although it is possible to find different ways to quantify the spin squeezing in the literature~\cite{Nori_rep11}, throughout this work we will use the spin squeezing parameter ($\xi$) proposed by Kitagawa and Ueda~\cite{Kitagawa_pra93}. In this approach, spin-squeezed states are quantum-correlated states with reduced fluctuations in one of the collective spin components, and quantum correlations are essential ingredients to characterize spin squeezing unambiguously. The suitable interaction among the spins can cancel out the fluctuation in one specific direction and reduce fluctuations in another~\cite{Kitagawa_pra93}.  In this regard, Kitagawa and Ueda's spin-squeezing parameter is given by 
\begin{equation}
\xi=\frac{\mathrm{4}\left(\Delta\mathcal{S}_{n_{\perp}}^{2}\right)_{\text{min}}}{N},\label{eq:2}
\end{equation}
where $\mathcal{S}_{\alpha}=\frac{1}{2}\sum_{i}^{N}\sigma_{i\alpha}, \,(\alpha=x,y,z)$ represents the collective spin operators of an ensemble of $N$ spin-$1/2$ particles (or qubits), $\sigma_{i\alpha}$ are the Pauli matrices for the \textit{i}th spin and $\vec{n}_{\perp}$ denotes an axis perpendicular net polarization of the collective spin, i.e., the mean
spin direction, $\vec{n}_{0}=\frac{\langle\mathcal{S}\rangle}{\vert\langle\mathcal{S}\rangle\vert}$~\cite{Kitagawa_pra93,Sanders_pra03}. In turn, a collective spin state can be viewed as a squeezed state if the variance of the spin normal component, $\Delta\mathcal{S}_{n_{\perp}}^{2}$, is less than the standard quantum limit $N/4$, which represents the variance associated with the coherent spin state~\cite{Kitagawa_pra93,Nori_rep11}. In other words, if the inequality $\xi<1$ is satisfied the system is spin-squeezed. It is worth noting that this way of quantifying the spin squeezing is independent of the coordinate system and highlights the role of quantum correlation in the notion of squeezing~\cite{Kitagawa_pra93}. In Appendix \ref{ApendA}, we outlined how to explicitly calculate the Eq. (\ref{eq:2}) for an initial state that has the mean spin direction pointing out in the $z$-direction. 

It is essential to stress that a nonlinear interaction is required to correlate each elementary spin and create a quantum correlation since a global linear interaction only rotates the individual spins and does not produce quantum correlations between them. As pointed out by Kitagawa and Ueda, it is possible to use two classes of nonlinear spin Hamiltonian, namely, as one-axis twisting: $\mathcal{H}_{\text{OAT}}=\kappa\mathcal{S}_{x}^{2}$ and two-axis countertwisting: $\mathcal{H}_{\text{TAT}}=\frac{\kappa}{2i}(\mathcal{S}_{+}^{2}-\mathcal{S}_{-}^{2})$ to produce spin squeezing effects. Both kinds of Hamiltonian have been widely explored theoretically~\cite{Kitagawa_pra93,Sanders_pra03,Ueda_pra14,Kajtoch_pra15} and experimentally~\cite{Auccaise_prl15,Guang_prl07,Fernholz_prl08,Takeuchi_prl05,Sinha_QI03} to generate, characterize, and analyze the dynamics of spin squeezing in different setups and platforms. In this work, we consider the so-called one-axis twisting nonlinear spin-squeezing interaction coupled with an external transversal field ($\hbar=1$)
\begin{equation}
\mathcal{H}=\kappa\mathcal{S}_{x}^{2}+\Omega\mathcal{S}_{z},
\label{eq2}
\end{equation}
where $\kappa\geq0$ describes the intensity of the nonlinear parameter associated with the spin squeezing and $\Omega\geq0$ represents the strength of the external field tuned in the $z$-direction~\cite{Nori_rep11,Sanders_pra03,Law_pra01}. This kind of Hamiltonian has been used to describe a four-level quantum heat engine in an Otto cycle with a working medium of two spins subject to a nonlinear interaction~\cite{Altintas_pre14} as well as useful in characterizing the dynamics of a Bose-Einstein condensate in a double-well potential~\cite{Milburn_pra97,Zoller_pra02,Jing_pla02}.

\section{Two-stroke thermal machine using spin squeezing operation}\label{secIII}

We consider a two-stroke quantum machine based on a nonlinear spin squeezing operation outlined in Eq. (\ref{eq2}). The two-stroke structure for the cycle has been considered in previous works \cite{Timpanaro2019, Sacchi2021, Altintas_pre14} with different interaction Hamiltonians. We highlight that the main difference when comparing our cycle with the previous ones is that we explicitly consider the role played by the coherence along the cycle as well as its relation with the spin squeezing. In particular, \cite{Altintas_pre14} also considers the Hamiltonian given by Eq. (\ref{eq2}), but in a quasi-static Otto cycle, which fundamentally differs from our finite-time two-stroke quantum cycle illustrated in Fig.~\ref{CycleIllustration} and engendered as follows:
\begin{figure}[hbt]
\includegraphics[scale=0.48]{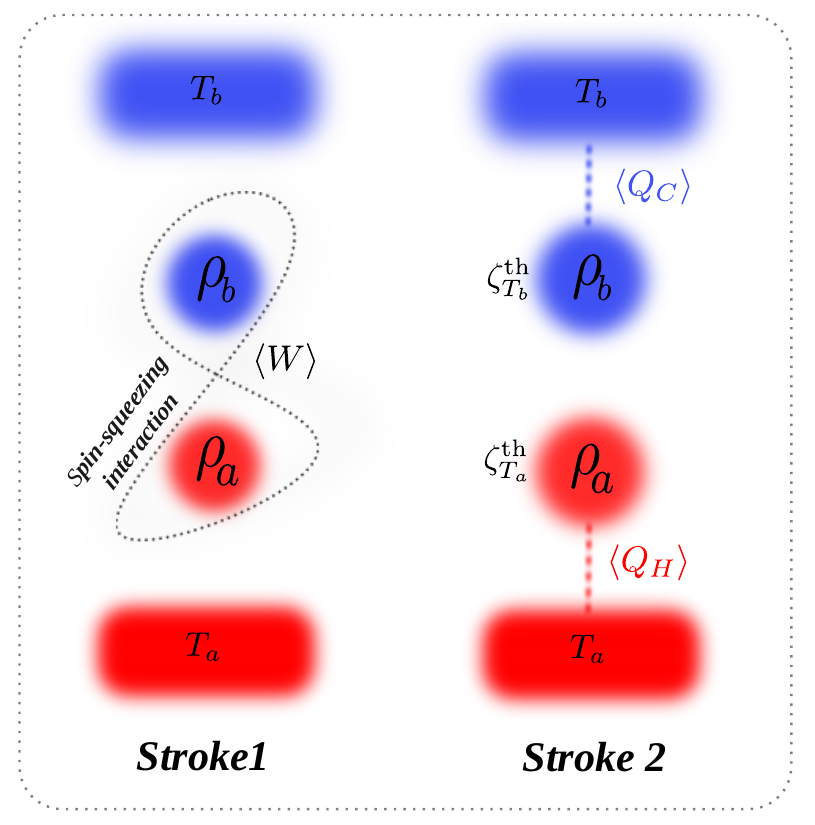}
\caption{Two-stroke thermal machine cycle. \textit{Stroke 1} - At the time $t_{0}$, we assume that the working medium is composed of two spins initially prepared in a state $\rho_{t_{0}}=\rho_{a}(t_{0})\otimes\rho_{b}(t_{0})$ where the reduced local states $\rho_{i}\left(t_{0}\right)=\zeta_{T_{i}}^{\text{th}}$
are thermal equilibrium states. Immediately afterward, during a period $\tau$,
a nonlinear spin squeezing operation, described by the Hamiltonian  $\mathcal{H}_{ab}=\kappa\mathcal{S}_{x}^{2}+\Omega\mathcal{S}_{z}$, is switch-on.  \textit{Stroke 2} - In order to close the two-stroke thermal machine cycle, the two qubits are allowed to interact with their respective thermal reservoirs until to reach the complete local thermalization.}
\label{CycleIllustration}
\end{figure}

\begin{figure}[hbt]
\includegraphics[scale=0.4]{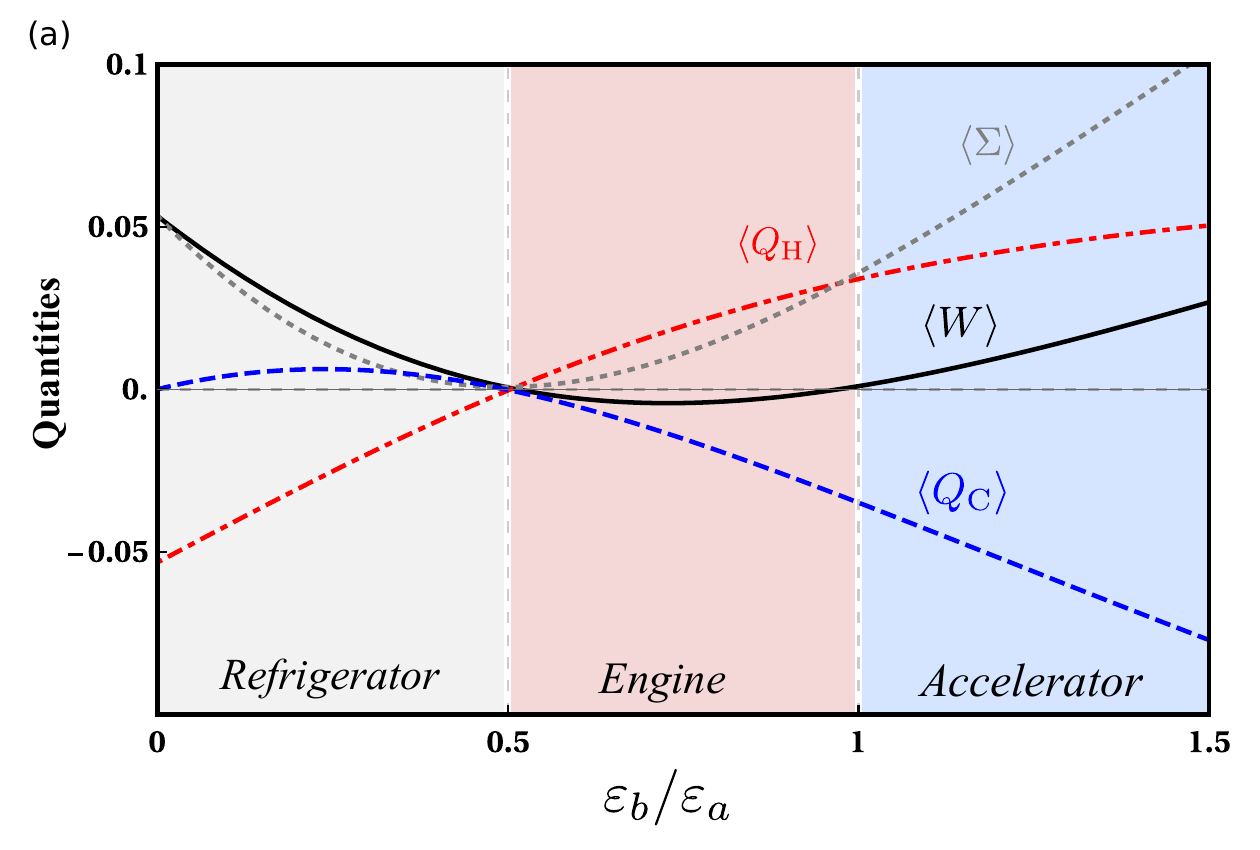}
\vspace{0.2cm}
\includegraphics[scale=0.4]{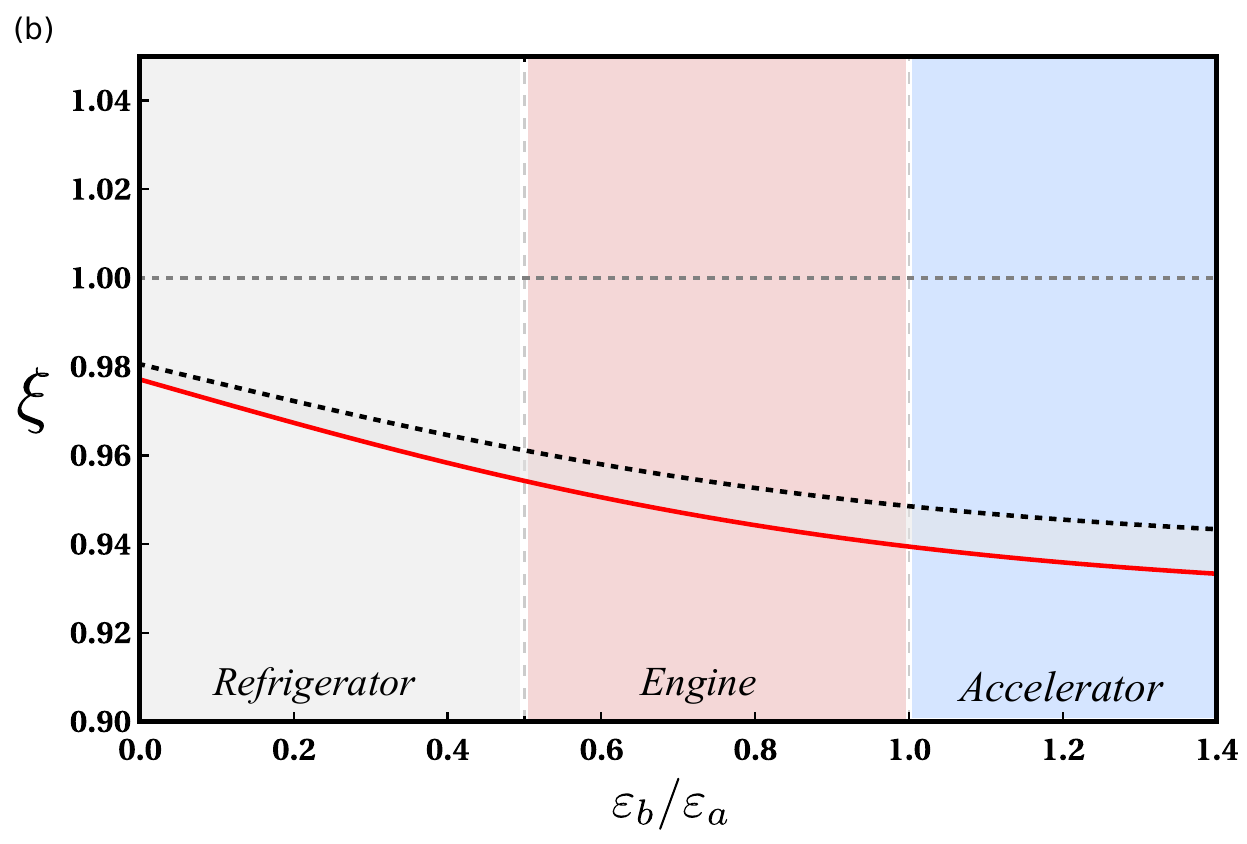}
\caption{(a) Thermodynamics quantities for the two-stroke quantum machine. Averages of the extracted work $\langle W\rangle$ (black solid line), released heat by the hot bath $\langle Q_{\text{H}}\rangle$ (red dash-dotted line), cold heat $\langle Q_{\text{C}}\rangle$ (blue dashed line) and entropy production $\langle\Sigma\rangle$ (gray dotted line) as a function of $\varepsilon_{b}/\varepsilon_{a}$. We considered the following set of parameters: $\varepsilon_{a}=1$, $\beta_{a}=1$, $\beta_{b}=2\beta_{a}$, $\kappa=1$, $\Omega=10\kappa$ and $\tau=\kappa$. (b) Spin squeezing parameter $\xi$ as a function of the ratio $\varepsilon_b/\varepsilon_a$ to explicitly show the similar spin squeezing behavior for the refrigerator, engine, and accelerator regimes. We have fixed the parameters to be $\Omega = 10\kappa$, $\tau=10\kappa$, $\varepsilon_{a} = 1$, $\beta_a = 1$, $\beta_b = 2\beta_a$, $\kappa = 0.1$ (black curve) and $\kappa = 0.12$ (red curve).}
\label{Fig01}
\end{figure}

\textit{\textbf{Stroke 1}} - At the time $t_{0}$, we assume that
the working medium is composed of two spins initially prepared in a state $\rho_{t_{0}}=\rho_{a}(t_{0})\otimes\rho_{b}(t_{0})$ where the reduced local states $\rho_{i}\left(t_{0}\right)=\zeta_{T_{i}}^{\text{th}}$
are thermal equilibrium states $\zeta_{T_{i}}^{\text{th}}=e^{-\beta_{i}\mathcal{H}_{i}}/\mathcal{Z}_{i}$ with  $\mathcal{Z}_{i}=\text{tr}[e^{-\beta_{i}\mathcal{H}_{i}}]$, $\beta_{i}=1/T_{i}$,  and $i=[a,b]$ (we set the Boltzmann's constant $k_B=1$). The local Hamiltonian of each qubit is given by $\mathcal{H}_{i}=-\varepsilon_{i}\sigma_{+}^{i}\sigma_{-}^{i}$, where $\sigma_{\pm}=(\sigma_{x}\pm i\sigma_{y})/2$ are the ladder operators. Here, we consider that the qubit (a) is hotter than the qubit (b), i.e., $T_{a}>T_{b}$.

Immediately afterward, during a period $\tau$,
a nonlinear spin squeezing operation, described by the Hamiltonian  $\mathcal{H}_{ab}=\kappa\mathcal{S}_{x}^{2}+\Omega\mathcal{S}_{z}$, is switch-on. During this process, the total Hamiltonian of the system changes from $\mathcal{H}(t_{0})=\mathcal{H}_{0}=\mathcal{H}_{a}\otimes\mathbb{I}_{b}+\mathbb{I}_{a}\otimes\mathcal{H}_{b}$ to $\mathcal{H}_\tau=\mathcal{H}_{0}+\mathcal{H}_{ab}$, creating a quantum correlated spin squeezed state $\rho_{\tau}=\mathcal{U}_{\tau}\rho_{0}\mathcal{U}^{\dagger}_{\tau}$. Here, $\mathbb{I}$ represents a two by two identity matrix for each subspace. Physically, the process described in stroke 1 is carried out by modifying a parameter in the system's Hamiltonian through an external agent. Thus, since at this time the working substance is completely isolated from its environment, the variation in its internal energy is entirely attributed to the work performed on the system, as dictated by the second law of thermodynamics.

It is worth mentioning that the driving evolution during this process 
is precisely described through the time evolution operator $\mathcal{U}_{\tau}=\exp(-i\mathcal{H}_{\tau}\tau)$. Nevertheless, to obtain all relevant thermodynamics quantities, especially in the engine regime which will be our focus here, the parameters used in our numerical simulations show that we can rule out the contribution from the free local Hamiltonian and effectively describe the unitary evolution by considering only the interaction Hamiltonian, resulting in $\mathcal{U}_{\tau}\approx\mathcal{U}_{ab}=\exp(-i\mathcal{H}_{ab}\tau)$. In Appendix~\ref{ApendB}, we present numerical simulations for both cases, considering and excluding the free Hamiltonian contribution in the time evolution operator, to reinforce that the behavior of the energetic quantities as well as the spin-squeezing parameter remains very similar.

\textit{\textbf{Stroke 2}} - In order to close the two-stroke thermal machine cycle, the two qubits are allowed to interact with their respective thermal reservoirs until to reach the complete local thermalization. In this stroke, the local energies exchanged are associated with heat, that will be released or absorbed to/from the local reservoirs depending on the set of parameters. 

For the two-stroke thermal machine, the relevant thermodynamics quantities are the average extracted or performed work due to the unitary interaction
\begin{equation}
\langle W\rangle=\text{Tr}[(\mathcal{H}_{a}\otimes\mathbb{I}_b+\mathbb{I}_a\otimes\mathcal{H}_{b})\rho_{\tau}]-\text{Tr}[(\mathcal{H}_{a}\otimes\mathbb{I}_b+\mathbb{I}_a\otimes\mathcal{H}_{b})\rho_{0}],\label{W}    
\end{equation}
with $\rho_{\tau}=\mathcal{U}_{\tau}\rho_{0}\mathcal{U}_{\tau}^{\dagger}$, the heat released or absorbed to the hot and cold thermal
reservoirs, respectively, 
\begin{equation}
\langle Q_{\text{H}}\rangle=-\text{Tr}[\mathcal{H}_{a}\otimes\mathbb{I}_b(\rho_{\tau}-\rho_{0})],\label{QH}
\end{equation}
\begin{equation}
\langle Q_{\text{C}}\rangle=-\text{Tr}[\mathbb{I}_a\otimes\mathcal{H}_{b}(\rho_{\tau}-\rho_{0})],\label{QC}
\end{equation}
and the total entropy production
\begin{equation}
\langle\Sigma\rangle =  (\beta_{b}-\beta_{a})\langle Q_{\text{H}}\rangle+\beta_{b}\langle Q_{\text{C}}\rangle\label{sigama1}. 
\end{equation}
For each operating regime, the following conditions must be satisfied. (i) heat engine: $\langle Q_{\text{H}}\rangle>0$, $\langle Q_{\text{C}}\rangle<0$,
and $\langle W\rangle<0$, (ii) refrigerator: $\langle Q_{\text{H}}\rangle<0$, $\langle Q_{\text{C}}\rangle>0$, and $\langle W\rangle>0$, and  (iii) accelerator: $\langle Q_{\text{H}}\rangle>0$, $\langle Q_{\text{C}}\rangle<0$, and $\langle W\rangle>0$. Remark that the entropy production is always positive, regardless of the cycle configuration, obeying the second law of thermodynamics $\langle\Sigma\rangle \geq 0$~\cite{Landi2021}.

Figure~\ref{Fig01}(a) illustrates the thermodynamic quantities for the three possible regimes of operation considering our two-stroke quantum thermal machine. As can be observed, by setting the appropriate ratio of the energy gaps between each qubit, $\epsilon_b/\epsilon_a$, the machine can operate as a refrigerator, engine, or accelerator. Just for completeness, in Fig.~\ref{Fig01}-(b) we show the behavior of the spin squeezing parameter $\xi$ as a function of the ratio $\varepsilon_b/\varepsilon_a$, covering all engine operating regimes of the two-stroke cycle. The primary purpose is to illustrate that the squeezing effect could also be exploited in other operational regimes as well. 

From this point forward, we restrict our analysis to the quantum engine regime of operation. To quantify the engine's performance and the nonequilibrium thermodynamic quantities, obtained straightforwardly in Appendix~\ref{ApendB} using Eqs.~(\ref{W}-\ref{sigama1}) and alternatively employing the characteristic function approach, we analyze two key figures of merit: efficiency, defined as $\eta=-\langle W\rangle/\langle Q_{\text{H}}\rangle$, and the extracted power, given by $\mathcal{P} = - \langle W \rangle /\tau$, where $\tau$ represents the interaction time (see details in Fig.~\ref{Effandpower}).

\begin{figure}
\includegraphics[scale=0.4]{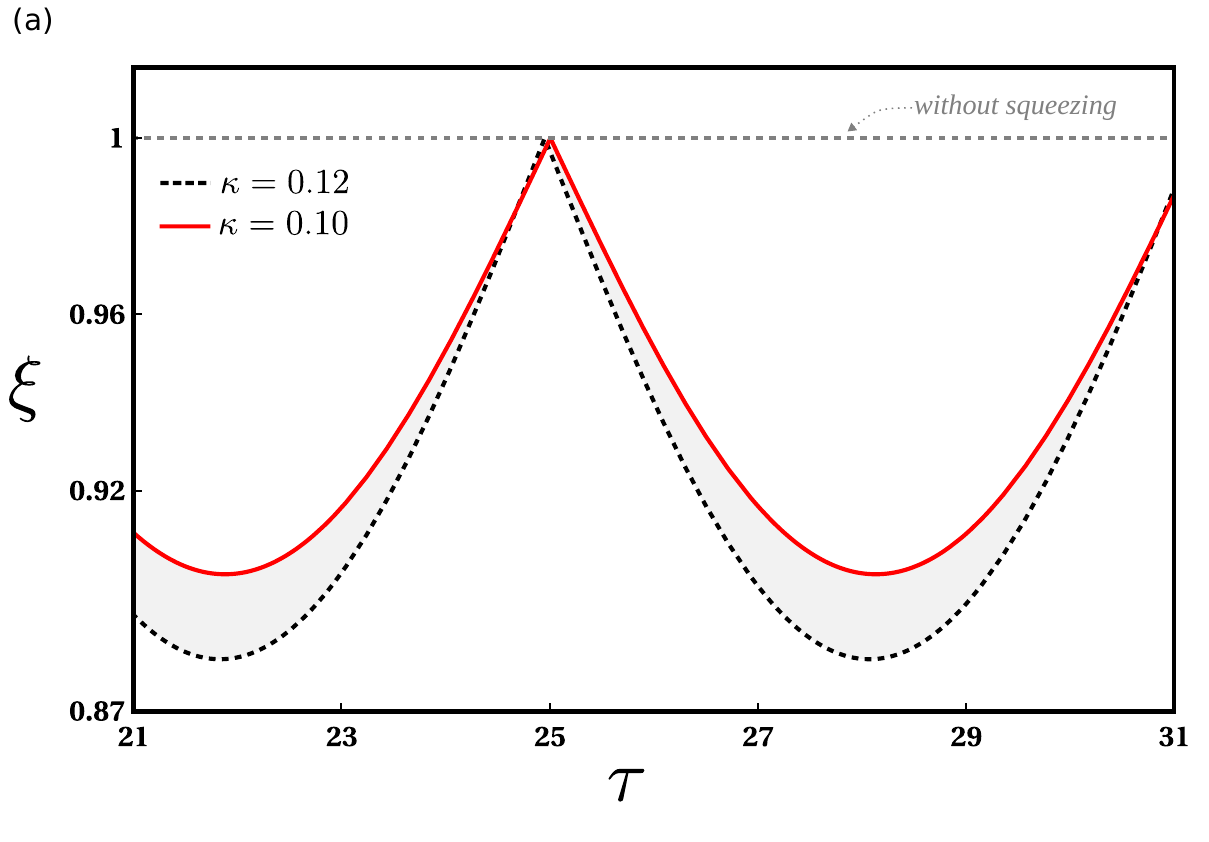}
\includegraphics[scale=0.4]{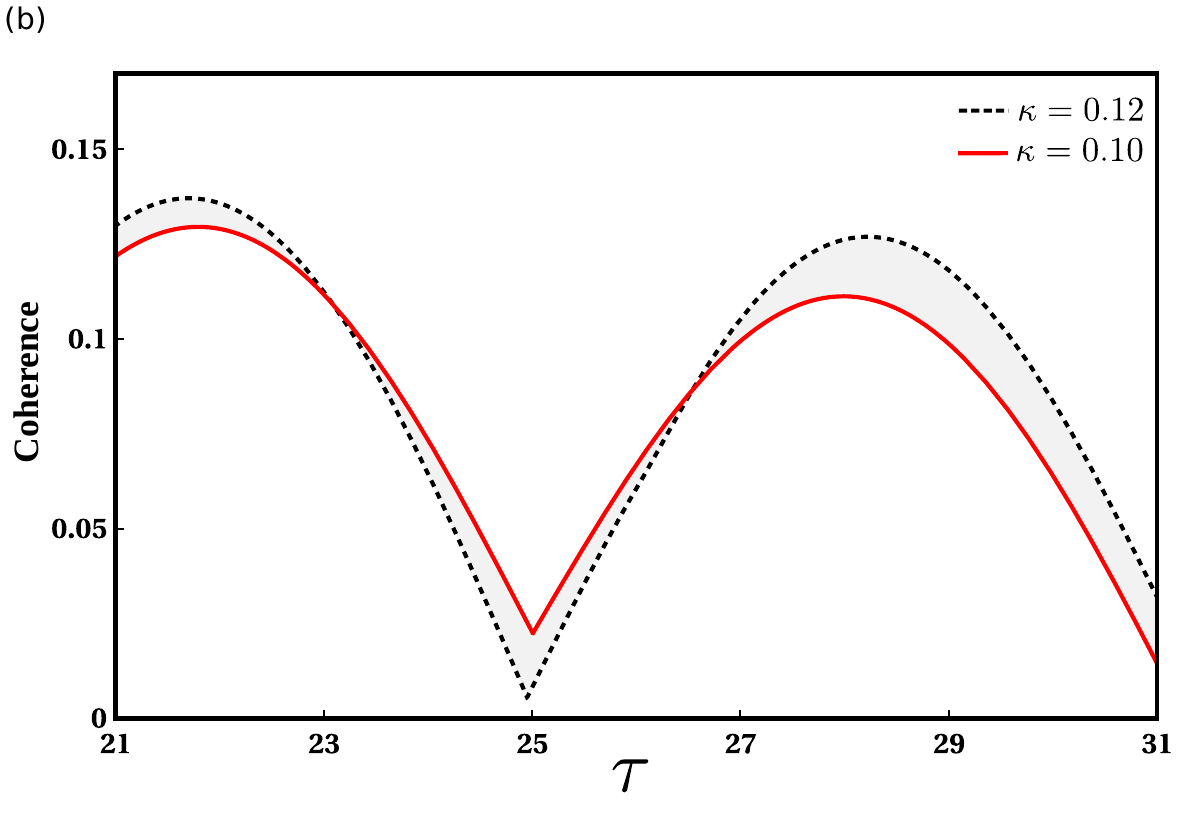}
\caption{(a) Spin squeezing parameter $\xi$ as a function of the interaction time for
different values of nonlinear parameter: $\kappa=0.12$ (black dashed line), $\kappa=0.10$
(red solid line). (b) $l_{1}$-norm of coherence as a function of $\tau$ to quantify the global coherence
in the energy basis. In both cases, we consider the following set of parameters:
$\beta_{a}=1$, $\beta_{b}=2\beta_{a}$, $\varepsilon_a=1$, $\varepsilon_b=0.6\varepsilon_a$ and $\Omega=0.5$.}
\label{Sque_Cohe}
\end{figure}

\begin{figure*}[t]
\includegraphics[scale=0.4]{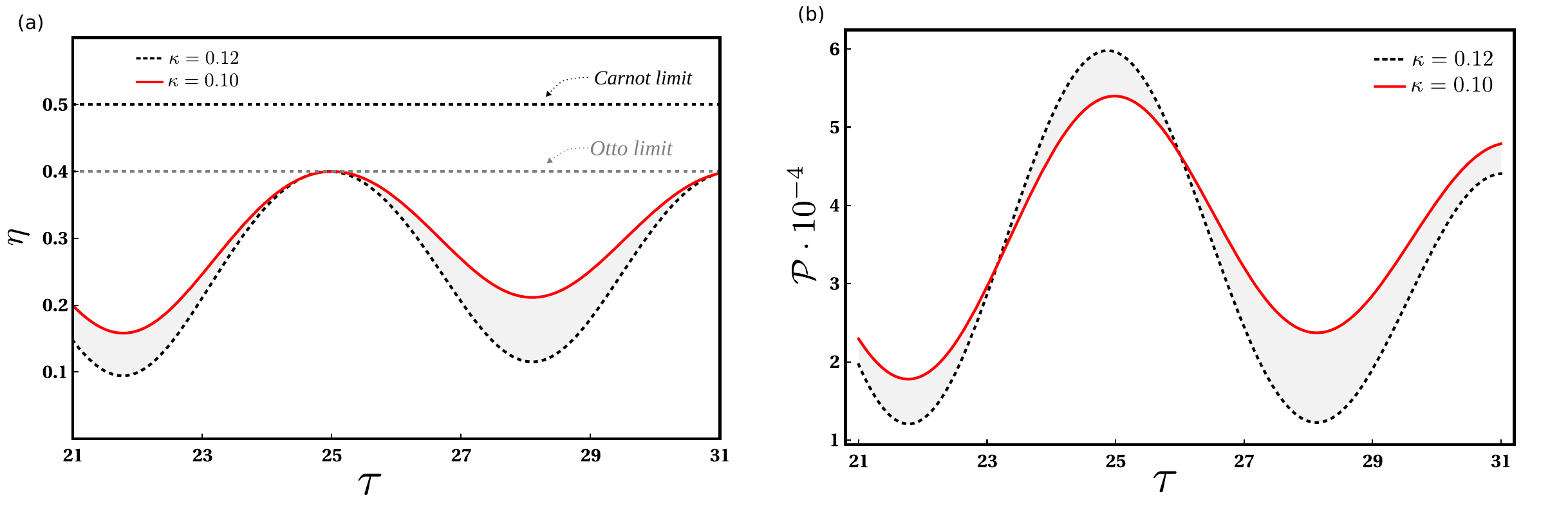}
\caption{(a) Efficiency and (b) extracted power of the two-stroke quantum heat engine as a function of the interaction time for
different values of nonlinear parameter: $\kappa=0.12$ (black dashed line), $\kappa=0.10$
(red solid line). We have considered the same set of parameters as in Fig. \ref{Sque_Cohe}. }
\label{Effandpower}
\end{figure*}

\section{Results}\label{secV}

We begin our investigation of the quantum engine by examining the spin squeezing generated during this mode of operation. From Eq. (\ref{eq2}) it is clear that the nonlinear feature of the engine is dictated by the intensity of the parameter
$\kappa$. With this in mind, Fig. \ref{Sque_Cohe}-(a) depicts
the Kitagawa and Ueda's spin squeezing parameter as a function of the interaction time for different values of nonlinear parameter $\kappa$, to show how the parameter $\xi$  depends on interaction time. We observe that due to the orthogonal contribution, encoded in the parameter $\Omega$, turned on during our analysis, the intensity of the spin squeezing in the engine regime is attenuated to some values of the interaction time. The choice in the specific range of time interaction $\tau$ is appropriated to ensure the engine regime as displayed in Fig.~\ref{Fig01}(a), considering dimensionless units.

Likewise, in Fig. \ref{Sque_Cohe}-(b), we show how spin squeezing is directly associated with the production of coherence
in the global basis of the system, which in turn, is a signature
of quantum correlation \cite{Baumgratz2014}. In Fig. \ref{Sque_Cohe}-(b) we show the
$l_{1}$-norm to quantify the dynamics of coherence in the global basis (energy basis)
of the two spins. Notice that, the coherence dynamics during the engine regime are directly linked to the spin squeezing produced by the nonlinear interaction. Specifically, the maximum values of the coherence match with the maximums of the spin squeezing parameter and vice versa. This highlights the role played by the
term $\kappa$ in the nonlinear Hamiltonian in Eq. (\ref{eq2}), by controlling
the amount of coherence generated in the first stroke. 

We now focus on the thermodynamics and performance of the two-stroke quantum heat engine and spin-squeezing effects. From the thermodynamic
quantities obtained in Appendix \ref{ApendB}) and choosing the appropriated relation between the two frequencies $\varepsilon_{a}$ and
$\varepsilon_{b}$ such that the cycle operates as an engine (see Fig. \ref{Fig01}(a)) the figures of merit as efficiency and extracted power can be analyzed.

Figure \ref{Effandpower}-(a) and \ref{Effandpower}-(b) illustrate the behavior of efficiency and extracted power as a function of the interaction time for two values of the nonlinear parameter $\kappa$, respectively. For clarity, in Fig. \ref{Effandpower}-(a) we also indicate the Otto and Carnot efficiencies for the parameters considered. The oscillatory behavior in both efficiency and extracted power is clearly due to the introduction of the spin squeezing effects in the system Hamiltonian, whereas the intensity of it depends on the value of $\kappa$. We observe that irrespective of the value of $\kappa$, it is possible to reach the Otto efficiency provided a sufficiently high control over the systems parameter, i.e., the intensity of the spin squeezing and the squeezing operation time. Figure \ref{Effandpower}-(b) illustrates the fact that the maximum value of the extracted power follows that of the efficiency. Both of them are in agreement with the minimum value of spin squeezing or coherence, see Fig \ref{Sque_Cohe}. 

\begin{figure}[!h]
\centering
\includegraphics[scale=0.4]{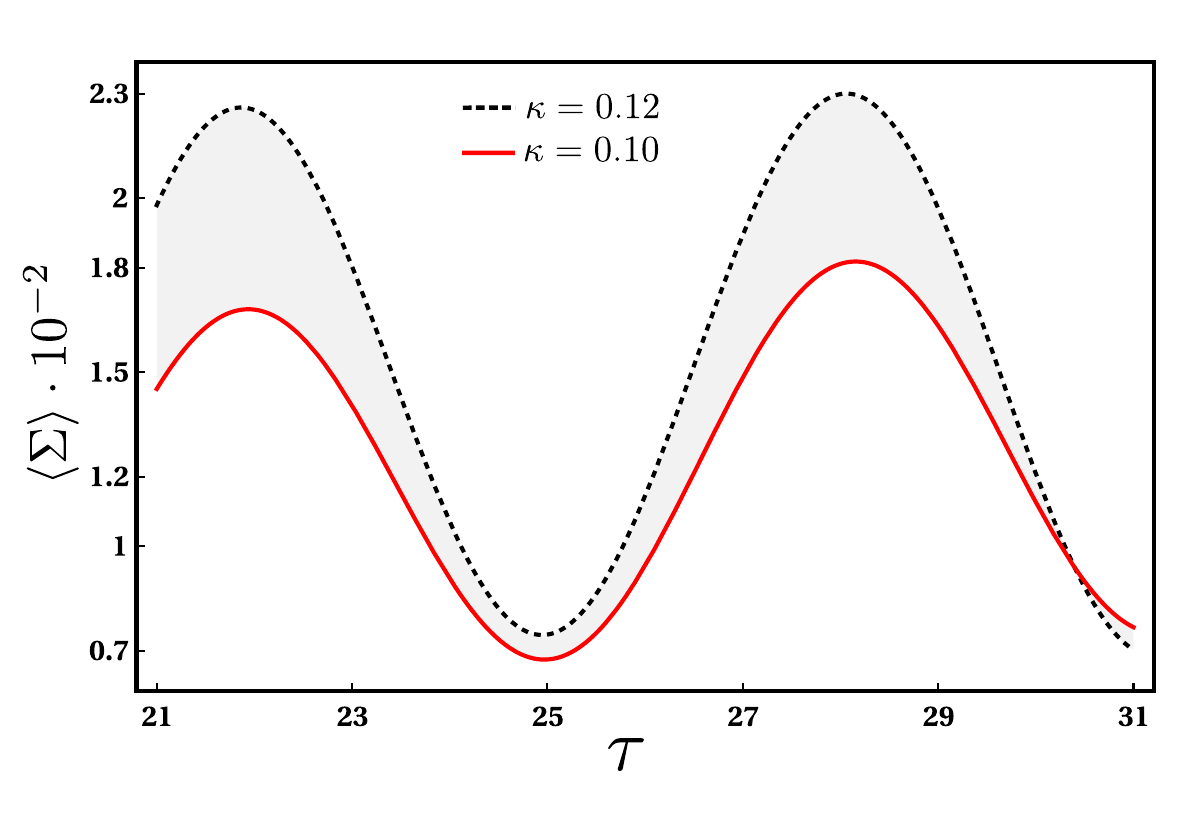}
\caption{Total entropy production as a function of the interaction time for
different values of nonlinear parameter: $\kappa=0.12$ (black dashed line), $\kappa=0.10$
(red solid line). We consider the same set of parameters displayed in Fig. \ref{Sque_Cohe}.}
\label{EPfigure}
\end{figure}

The total entropy production $\langle\Sigma\rangle$ along the cycle is also a relevant thermodynamic quantity once it amounts to the degree of irreversibility in a general process or cycle \cite{Landi2021,Santos2019}. By using the characteristic function we can express exactly the total entropy production for the engine cycle as in Eq.~(\ref{qhappb}). Figure \ref{EPfigure} depicts the total entropy production for the engine cycle as a function of the interaction time for two values of $\kappa$. Again, the spin squeezing effect introduces the oscillatory behavior and we observe that the minimum entropy production value coincides with the maximum efficiency for the cycle, in agreement with the general result in Ref. \cite{Camati2019}. We highlight that a suitable value for the total entropy production may be achieved provided a high control over the experimental parameters.

\section{Conclusion}\label{secVI}

We have theoretically investigated a quantum thermal machine model where spin-squeezing interaction is intrinsically considered in the system. First, we verified that by changing the energy gap ratio between the two qubits it is possible to operate the cycle in a refrigerator, engine, or accelerator regime. We have chosen the engine configuration to study the relation between the performance and spin squeezing degree. To quantify the degree of spin squeezing during the cycle dynamics, we employed Kitagawa and Ueda's parameter and we show that the coherent amount is linked to the spin squeezing parameter. We stress the fact that for other regimes of operation, such as refrigerator and accelerator, the aspects concerning the spin squeezing and coherence are similar, the only modification being the amount of each quantity.

All thermodynamic quantities to characterize the engine performance were evaluated through the characteristic formalism. The performance of the engine was studied through the efficiency and extracted power. We verified that the more the amount of spin squeezing in a given time, the less the efficiency and extracted power. Thus, for a cycle in which the spin squeezing operation is always on, the best performance may be achieved provided a high control over the parameters. We also considered the irreversibility of the cycle by computing the total entropy production. The results indicate that the more the spin squeezing degree, the more the total entropy production in a cycle. This aspect is associated with the performance of the engine for a given value of the interaction time. We hope that this work can contribute to unveiling the role played by quantum correlations in nontrivial quantum thermal machines. Finally, our model also allows for experimental implementation, for instance, in nuclear magnetic resonance.


\section*{Acknowledgments}
Carlos H. S. Vieira acknowledges the São Paulo Research Foundation (FAPESP), Grant. No. 2023/13362-0, for financial support and the Federal University of ABC (UFABC) to provide the workspace. Jonas F. G. Santos acknowledges CNPq Grant No. 420549/2023-4, Fundect and Universidade Federal da Grande Dourados for support.



\newpage

\appendix
\begin{widetext}
\section{Spin squeezing parameter and nonlinear evolution}\label{ApendA}


\textbf{Kitagawa and Ueda’s parameter} - In this section, we explicit the Eq.~(\ref{eq:2})
for a quantum initial state which has the mean spin direction (MSD), defined by
$\vec{n}_{0}$ in the $z$ direction
\begin{equation}
\vec{n}_{0}=\frac{\langle{\mathcal{S}}\rangle}{|\langle\mathcal{S}\rangle|}=\frac{\mathcal{S}_{z}}{|\langle\mathcal{S}\rangle|}.\label{ap1}
\end{equation}
After the interaction with one kind of nonlinear Hamiltonian the final
state, $\rho_{\tau}=\mathcal{U}_\tau\rho_{0}\mathcal{U}^{\dagger}_\tau$,
may have its fluctuation reduced in some direction. Since the MSD is in the $z$ direction, we can introduce an orthogonal vector as $\vec{n}_{\perp}=\hat{x}\cos(\phi)+\hat{y}\sin(\phi)$
and therefore the spin component becomes
\begin{equation}
\mathcal{S}_{n_{\perp}}=\mathcal{S}_{x}\cos(\phi)+\mathcal{S}_{y}\sin(\phi),\label{ap2}
\end{equation}
where $\phi$ is an arbitrary angle belonging to the $xy$ plane. In turn, for this particular initial state, we have that the variance is equal to $\Delta\mathcal{S}_{n_{\perp}}^{2}=\langle \mathcal{S}_{n_{\perp}}^{2}\rangle$,
since the average are $\langle\mathcal{S}_{x}\rangle=\langle\mathcal{S}_{y}\rangle=0$.
Thereby, the variance of the spin component in the orthogonal direction for the final state $\rho_{\tau}$ becomes
\begin{align}
\left(\Delta\mathcal{S}_{n_{\perp}}^{2}\right) & =\frac{1}{2}\langle \mathcal{S}_{x}^{2}+\mathcal{S}_{y}^{2}\rangle+\frac{1}{2}\langle \mathcal{S}_{x}^{2}-\mathcal{S}_{y}^{2}\rangle\cos(2\phi) +\frac{1}{2}\sin(2\phi)\langle\left\{ \mathcal{S}_{x},\mathcal{S}_{y}\right\} \rangle,\label{ap3}
\end{align}
where we used the anti-commutator term $\left\{ \mathcal{S}_{x},\mathcal{S}_{y}\right\} =\mathcal{S}_{x}\mathcal{S}_{y}+\mathcal{S}_{y}\mathcal{S}_{x}$
and trigonometric relations. As shown in Eq. (\ref{ap3}), the variance $\Delta\mathcal{S}_{n_{\perp}}^{2}$ is a function of the angle $\phi$ and therefore its minimization can be obtained through the derivation of the equation concerning this angle providing the following reduced variance
\begin{equation}
\delta_{-}=\frac{1}{2}\left[\langle \mathcal{S}_{x}^{2}+\mathcal{S}_{y}^{2}\rangle-\sqrt{\langle \mathcal{S}_{x}^{2}-\mathcal{S}_{y}^{2}\rangle^{2}+\langle\left\{ \mathcal{S}_{x},\mathcal{S}_{y}\right\} \rangle^{2}}\right],\label{ap4}
\end{equation}
where $\delta_{-}\equiv\left(\Delta\mathcal{S}_{n_{\perp}}^{2}\right)_\text{min}$
corresponding the squeezing along the orthogonal direction $\vec{n}_{\perp}$
with the respective optimally squeezing angle 
\begin{equation}
\phi_{\text{opt}}=\frac{\pi}{2}+\frac{1}{2}\arctan\left[\frac{\langle\left\{ \mathcal{S}_{x},\mathcal{S}_{y}\right\} \rangle^{2}}{\langle \mathcal{S}_{x}^{2}-\mathcal{S}_{y}^{2}\rangle}\right].\label{ap5}
\end{equation}

Therefore, using the Eq.~(\ref{ap4}), the Kitagawa and Ueda's
parameter, Eq.~(\ref{eq:2}), for an initial state, $\rho_0$, belonging to the
$z$-direction becomes
\begin{equation}
\xi=\alpha_{1}-\sqrt{\alpha_{2}+\alpha_{3}},\label{ap6}
\end{equation}
with,
\begin{equation}
\alpha_{1}=\frac{2\langle\mathcal{S}_{x}^{2}+\mathcal{S}_{y}^{2}\rangle}{N},\quad \alpha_{2}=\frac{4\langle\mathcal{S}_{x}^{2}-\mathcal{S}_{y}^{2}\rangle^{2}}{N^{2}},\quad \alpha_{3}=\frac{4\langle\left\{ \mathcal{S}_{x},\mathcal{S}_{y}\right\} \rangle^{2}}{N^{2}}.
\label{ap6}
\end{equation}
Although we do not explicitly, all quantities in Eqs.~(\ref{ap6}) are functions of the interaction time $\tau$ and the squeezing parameter 
$\kappa$, since all averages are obtained from the squeezed spin state $\rho_{\tau}=\mathcal{U}_{\tau}\rho_{0}\mathcal{U}_{\tau}^{\dagger}$.

\textbf{Nonlinear evolution} - Strictly, since the condition $[\mathcal{H}_{a}\otimes\mathbb{I}_b+\mathbb{I}_a\otimes\mathcal{H}_{b},\mathcal{H}_{ab}]\neq0$ holds for all nonzero $\kappa$, the nonlinear evolution to generate the spin-squeezing state, $\rho_{\tau}=\mathcal{U}_{\tau}\left(\zeta_{T_1}^{th}\otimes\zeta_{T_2}^{th}\right)\mathcal{U}_{\tau}^{\dagger}$, in the stroke 1 (see Fig.1), must be $\mathcal{U}_{\tau}=\exp\left[-i\tau\left(\mathcal{H}_{a}\otimes\mathbb{I}_b+\mathbb{I}_a\otimes\mathcal{H}_{b}+\mathcal{H}_{ab}\right)\right]$, where  $\mathcal{H}_{0}=\mathcal{H}_{a}\otimes\mathbb{I}_b+\mathbb{I}_a\otimes\mathcal{H}_{b}$ are the local free Hamiltonians and $\mathcal{H}_{ab}=\kappa\mathcal{S}_{x}^{2}+\Omega\mathcal{S}_{z}$ is the nonlinear spin-squeezing interaction. After some cumbersome calculations, it is possible to show that the unitary evolution $\mathcal{U}_{\tau}$ can be cast in the form:
\begin{equation}
\mathcal{U}_{\tau}=e^{\frac{-i\kappa\tau}{2}}\left(\begin{array}{cccc}
\Phi_{+} & 0 & 0 & \mu_{1}\sin\left(\frac{\gamma_{1}\tau}{2}\right)\\
0 & \Lambda_{+} & -i\mu_{2}\sin\left(\frac{\gamma_{2}\tau}{2}\right) & 0\\
0 & -i\mu_{2}\sin\left(\frac{\gamma_{2}\tau}{2}\right) & \Lambda_{-} & 0\\
\mu_{1}\sin\left(\frac{\gamma_{1}\tau}{2}\right) & 0 & 0 & \Phi_{-}
\end{array}\right)
\label{eq:21}
\end{equation}
with
\begin{equation}
\Phi_{\pm}=e^{\frac{i\varepsilon_{p}\tau}{2}}\left[\Theta_{\pm}\pm\frac{i\varepsilon_{p}}{\gamma_{1}}\sin\left(\frac{\gamma_{1}\tau}{2}\right)\right],\Lambda_{\pm}=e^{\frac{i\varepsilon_{p}\tau}{2}}\left[\cos\left(\frac{\gamma_{2}\tau}{2}\right)\pm\frac{i\Delta\varepsilon}{\gamma_{2}}\sin\left(\frac{\gamma_{2}\tau}{2}\right)\right],
\end{equation}
\begin{equation}
\quad\mu_{0}=-i\kappa\gamma_{0}^{-1},\quad\mu_{1}=-i\kappa e^{\frac{i\varepsilon_{p}\tau}{2}}\gamma_{1}^{-1},\quad\mu_{2}=\kappa e^{\frac{i\varepsilon_{p}\tau}{2}}\gamma_{2}^{-1},
\end{equation}
\begin{equation}
\gamma_{0}=\sqrt{\kappa^{2}+\Omega^{2}},\quad\gamma_{1}=\sqrt{\kappa^{2}+(\Omega-\varepsilon_{p}/2)^{2}},\quad\gamma_{2}=\sqrt{\kappa^{2}+(\Delta\varepsilon)^{2}},
\end{equation}
\begin{equation}
\Delta\varepsilon=\varepsilon_{a}-\varepsilon_{b},\quad \varepsilon_{p}=\varepsilon_{a}+\varepsilon_{b}.
\end{equation}
On the other hand, the unitary evolution considering only the interaction Hamiltonian $\mathcal{U}_{ab}=\exp\left(-i\mathcal{H}_{ab}\tau\right)$ yields

\begin{equation}
\mathcal{U}_{ab}=e^{\frac{-i\kappa\tau}{2}}\left(\begin{array}{cccc}
\Theta_{+} & 0 & 0 & \mu_{0}\sin\left(\frac{\gamma_{0}\tau}{2}\right)\\
0 & \cos\left(\frac{\kappa\tau}{2}\right) & -i\sin\left(\frac{\kappa\tau}{2}\right) & 0\\
0 & -i\sin\left(\frac{\kappa\tau}{2}\right) & \cos\left(\frac{\kappa\tau}{2}\right) & 0\\
\mu_{0}\sin\left(\frac{\gamma_{0}\tau}{2}\right) & 0 & 0 & \Theta_{-}
\end{array}\right),
\label{eq:20}
\end{equation}
where
\begin{equation}
\gamma_{0}=\sqrt{\kappa^{2}+\Omega^{2}}.
\end{equation}
Note that the unitary evolution shown in Eq. (\ref{eq:20}) can be directly obtained from Eq. (\ref{eq:21}) setting the conditions $\Delta\varepsilon\rightarrow 0$ and $\varepsilon_{p}\rightarrow 0$, i.e, $\mathcal{U}_{ab}=\mathcal{U}_{\tau}(\Delta\varepsilon\rightarrow0,\varepsilon_{p}\rightarrow0)$.

Henceforward, for simplicity, we explicitly all relevant quantities obtained from the squeezed spin state $\rho_{\tau}$, considering only the nonlinear evolution $\mathcal{U}_{ab}$. Additionally, for the chosen parameters set, our numeric simulations reinforce that the behavior of the spin-squeezing parameter (see Fig.~\ref{Figure09}) as well as the thermodynamic quantities (see Fig.~\ref{Figure10}) remains very similar. Therefore, the squeezed spin state $\rho_{\tau}=\mathcal{U}_{ab}\rho_{0}\mathcal{U}_{ab}^{\dagger}$ can be obtained, and the Kitagawa and Ueda's parameter, Eq.~(\ref{ap6}), for our two-stroke heat engine takes the following form: 
\begin{align}
 \xi(\kappa,\Omega,\tau)=1-\kappa\frac{\bar{z}\eta}{\gamma_{0}^{2}}\big\vert\sin\left(\frac{\gamma_{0}\tau}{2}\right)\big\vert,
 \label{spin_squez_our}
\end{align}
with
\begin{equation}
\bar{z}=1-\frac{1}{\mathcal{Z}_{b}}-\frac{1}{\mathcal{Z}_{a}},
\end{equation}
\begin{equation}
\eta=\sqrt{16\Omega^{2}+2\kappa^{2}[1+\cos(\gamma_{1} \tau)]}.
\end{equation}
It is easy to see from Eq.~(\ref{spin_squez_our}) that when there is not the nonlinear parameter, i.e., $\kappa=0$, there is no spin squeezing, i.e., $\xi(\kappa,\Omega,\tau)=1$. 
\begin{figure}[!t]
\centering
\includegraphics[width=0.4\linewidth]{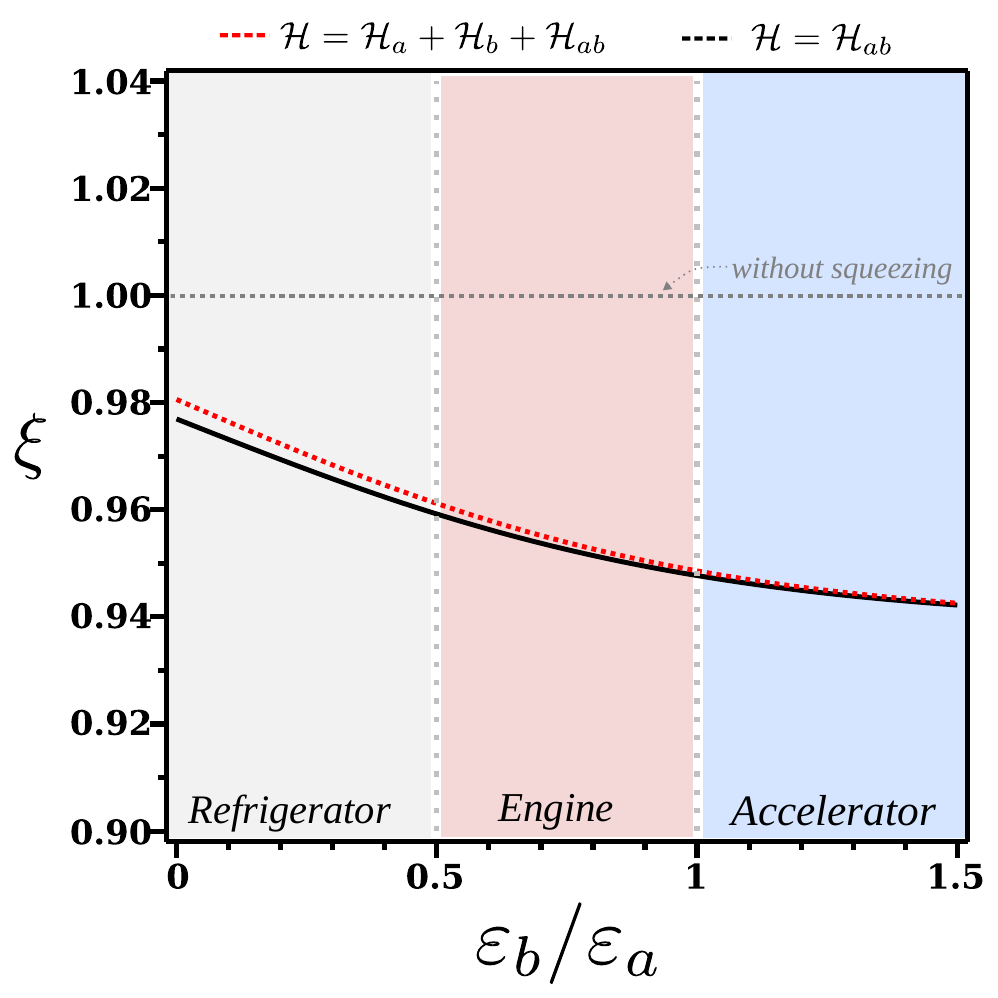}
\caption{The spin-squeezing parameter as a function of the ratio $\varepsilon_b/\varepsilon_a$ during the two-stroke heat engine. In the red dashed line, we consider the contribution of the free Hamiltonian $\mathcal{H}_{0}$ in the time evolution propagator or not (solid black line). We considered the same set of parameters to the main manuscript: $\varepsilon_{a}=1$, $\beta_{a}=1$, $\beta_{b}=2\beta_{a}$, $\kappa=0.1$, $\Omega=10\kappa$, and $\tau=\kappa$.}
\label{Figure09}
\end{figure}

\newpage
\section{Thermodynamics quantities and characteristic function}\label{ApendB}

Since the nonlinear unitary evolution, $\mathcal{U}_{ab}$ and the state squeezed state  $\rho_{\tau}=\mathcal{U}_{ab}\rho_{0}\mathcal{U}^{\dagger}_{as}$, can be derived as demonstrated in Appendix A, the thermodynamic quantities shown in the main manuscript, i.e., Eqs. (\ref{W}-\ref{sigama1}), assumes the following form
\begin{align}
\langle W\rangle&=\frac{\kappa^{2}(\varepsilon_{a}+\varepsilon_{b})}{\gamma^{2}\mathcal{Z}_{a}\mathcal{Z}_{b}}\sin^{2}\left(\gamma\tau/2\right)\left(e^{\beta_{a}\varepsilon_{a}+\beta_{b}\varepsilon_{b}}-1\right)+\frac{\left(1-\cos\left(\kappa\tau\right)\right)(\varepsilon_{a}-\varepsilon_{b})}{2\mathcal{Z}_{a}\mathcal{Z}_{b}}\left(e^{\beta_{a}\varepsilon_{a}}-e^{\beta_{b}\varepsilon_{b}}\right), \nonumber\\ \nonumber&
\langle Q_{\mathrm{H}}\rangle=\frac{\kappa^{2}\varepsilon_{a}}{\gamma^{2}\mathcal{Z}_{a}\mathcal{Z}_{b}}\sin^{2}\left(\gamma\tau/2\right)\left(1-e^{\beta_{a}\varepsilon_{a}+\beta_{b}\varepsilon_{b}}\right)-\frac{\varepsilon_{a}\left(1-\cos\left(\kappa\tau\right)\right)}{2\mathcal{Z}_{a}\mathcal{Z}_{b}}\left(e^{\beta_{a}\varepsilon_{a}}-e^{\beta_{b}\varepsilon_{b}}\right),\\ \nonumber&
\langle Q_{\mathrm{H}}\rangle=\frac{\kappa^{2}\varepsilon_{a}}{\gamma^{2}\mathcal{Z}_{a}\mathcal{Z}_{b}}\sin^{2}\left(\gamma\tau/2\right)\left(1-e^{\beta_{a}\varepsilon_{a}+\beta_{b}\varepsilon_{b}}\right)-\frac{\varepsilon_{a}\left(1-\cos\left(\kappa\tau\right)\right)}{2\mathcal{Z}_{a}\mathcal{Z}_{b}}\left(e^{\beta_{a}\varepsilon_{a}}-e^{\beta_{b}\varepsilon_{b}}\right),\\&
\langle\Sigma\rangle=\frac{\kappa^{2}\sin^{2}(\gamma\tau/2)(\beta_{a}\varepsilon_{a}+\beta_{b}\varepsilon_{b})}{\gamma^{2}\mathcal{Z}_{a}\mathcal{Z}_{b}}\left(e^{\beta_{a}\varepsilon_{a}+\beta_{b}\varepsilon_{b}}-1\right)+\frac{\left(1-\cos(\kappa\tau)\right)(\beta_{a}\varepsilon_{a}-\beta_{b}\varepsilon_{b})}{2\mathcal{Z}_{a}\mathcal{Z}_{b}}\left(e^{\beta_{a}\varepsilon_{a}}-e^{\beta_{b}\varepsilon_{b}}\right).
\label{qhappb}
\end{align}

Note that the nonequilibrium thermodynamic quantities are strongly linked to the $\kappa$, which quantifies the intensity of the nonlinearity associated with the spin-squeezing effects. Furthermore, it is important to stress that although would be possible to set $\Omega =0$ in the nonlinear Hamiltonian $\mathcal{H}_{ab}$, our numerical simulations evidence that it is impossible to implement a two-stroke quantum machine working in the heat engine regime which is our main focus in this work. 

In Fig.~\ref{Figure10}, we present numeric simulations for both cases, considering and excluding the free Hamiltonian contribution in the time evolution propagator, to reinforce that the behavior of the energetic quantities remains very similar during all two-stroke machine operations.

\begin{figure*}[t]
\centering
\includegraphics[scale=0.4]{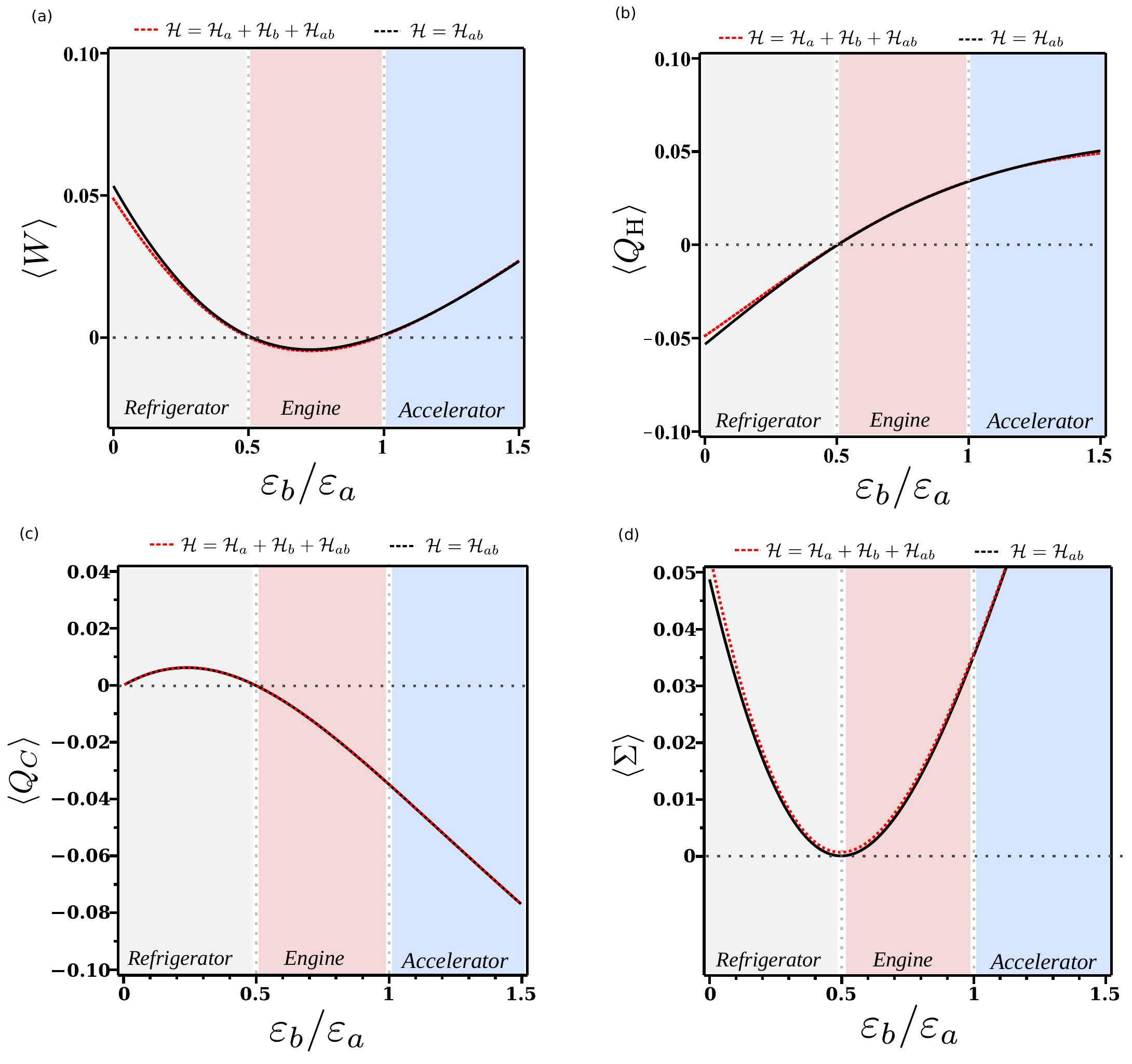}
\caption{Nonequilibrium thermodynamic quantities during the two-stroke quantum engine considering (dashed red line) and excluding (black solid line) the free Hamiltonian in the unitary evolution. (a) Averages of the extracted work $\langle W\rangle$, (b) released heat by the hot bath $\langle Q_{\text{H}}\rangle$, (c) cold heat $\langle Q_{\text{C}}\rangle$, and (d) entropy production $\langle\Sigma\rangle$ as a function of $\varepsilon_{b}/\varepsilon_{a}$. We considered the same set of parameters used in Fig.~\ref{Fig01} of the main manuscript.}
\label{Figure10}
\end{figure*}

\newpage
\textbf{Characteristic function approach} - An alternative way to obtain thermodynamic quantities along the two-stroke engine can be done by employing the characteristic function formalism. According to the two-point measurement protocol, the joint probability $P(W,Q_{H})$ associated with the work and heat can be derived using the correspondent characteristic function~\cite{Esposito_prl10,Sacchi2021}
\begin{align}\label{CF}
\mathcal{F}(\lambda,\nu)=\text{Tr}\left[\mathcal{U}_{\tau}^{\dagger}e^{i(\lambda-\nu)\mathcal{H}_{a}}e^{i\lambda\mathcal{H}_{b}}\mathcal{U}_{\tau}e^{-i(\lambda-\nu)\mathcal{H}_{a}}e^{-i\lambda\mathcal{H}_{b}}\rho_{0}\right],
\end{align}
where $\lambda$ and $\mu$ are the conjugate variables related with $W$ and $Q_{H}$, respectively.
Applying the Eq.(\ref{CF}), we can obtain the thermodynamic quantity associated with our two-stroke thermal machine as follows:
\begin{eqnarray}
\langle W^{n}Q_{\text{H}}^{m}\rangle=(-i)^{n+m}\frac{\partial^{n+m}\mathcal{G}(\lambda,\nu)}{\partial^{n}\lambda\partial^{m}\nu}\rvert_{_{\lambda=0,\nu=0}},
\label{WnQh}
\end{eqnarray} 
\begin{equation}
\langle\Sigma\rangle=(\beta_{b}-\beta_{a})\langle Q_{\text{H}}\rangle+\beta_{b}\langle W \rangle,
\label{sigma}
\end{equation}
\begin{equation}
\langle Q_{\text{C}}\rangle=-\langle W\rangle-\langle Q_{\text{H}}\rangle.
\label{Qc}
\end{equation}

In turn, using the individual Hamiltonians for each qubit, $\ensuremath{\mathcal{H}_{i}}=-\varepsilon_{i}\sigma_{+}^{i}\sigma_{-}^{i}$ where $i\in\{a,b\}$, the unitary evolution $\mathcal{U}_{ab}=\exp(-i\mathcal{H}_{ab}\tau)$, and the initial state, $\rho_0=\zeta_{T_1}^{th}\otimes\zeta_{T_2}^{th}$, the characteristics function becomes 
\begin{align}
\mathcal{F}(\lambda,\nu) & =\frac{1}{\mathcal{Z}_{a}\mathcal{Z}_{b}}\vert\Theta_{+}|^{2}\left(1+e^{\beta_{a}\varepsilon_{a}+\beta_{b}\varepsilon_{b}}\right)+\frac{1}{\mathcal{Z}_{a}\mathcal{Z}_{b}}\vert\cos\left(\frac{\kappa\tau}{2}\right)|^{2}\left(e^{\beta_{a}\varepsilon_{a}}+e^{\beta_{b}\varepsilon_{b}}\right)\nonumber \\
 & +\frac{1}{\mathcal{Z}_{a}\mathcal{Z}_{b}}\vert\mu_{0}\sin\left(\frac{\gamma_{0}\tau}{2}\right)|^{2}e^{-i\varepsilon_{a}(\lambda-\mu)-i\lambda\varepsilon_{b}}+\frac{1}{\mathcal{Z}_{a}\mathcal{Z}_{b}}\vert\mu_{0}\sin\left(\frac{\gamma_{0}\tau}{2}\right)|^{2}e^{i\varepsilon_{a}(\lambda-\mu)+i\lambda\varepsilon_{b}}e^{\beta_{a}\varepsilon_{a}+\beta_{b}\varepsilon_{b}}  \nonumber \\+
 &\frac{1}{\mathcal{Z}_{a}\mathcal{Z}_{b}}\vert\sin\left(\frac{\kappa\tau}{2}\right)|^{2}e^{i\varepsilon_{a}(\lambda-\mu)-i\lambda\varepsilon_{b}}e^{\beta_{a}\varepsilon_{a}}+\frac{1}{\mathcal{Z}_{a}\mathcal{Z}_{b}}\vert\sin\left(\frac{\kappa\tau}{2}\right)|^{2}e^{-i\varepsilon_{a}(\lambda-\mu)+i\lambda\varepsilon_{b}}e^{\beta_{b}\varepsilon_{b}}.\label{eq:22}
\end{align}

Thus, using Eq. (\ref{WnQh}-\ref{Qc}) it is possible to verify that the same set of nonequilibrium thermodynamic quantities is shown in Eqs. (\ref{qhappb}) are found.

\end{widetext}

\end{document}